\documentclass[a4paper,11pt]{article}
\pdfoutput=1 

\usepackage{jcappub} 

\usepackage[T1]{fontenc} 
\usepackage{graphicx}  
\usepackage{dcolumn}   
\usepackage{bm}        
\usepackage{amssymb}   
\usepackage{mathrsfs}  
\usepackage{amsmath}  
\usepackage{gensymb} 
\usepackage[caption=false]{subfig} 
\usepackage{hyperref}

\title{\boldmath Sensitivity of the CUORE detector to $14.4$ keV solar axions emitted by the M1 nuclear transition of$~^{57}$Fe}

\author[a,1]{Dawei Li,\note{Corresponding author.}}
\author[a]{Richard J. Creswick,}
\author[a]{Frank T. Avignone III,}
\author[b]{and Yuanxu Wang}

\affiliation[a]{Department of Physics and Astronomy, University of South Carolina, Columbia, SC, USA}
\affiliation[b]{School of Physics and Electronics, Henan University, Kaifeng, Henan, China}

\emailAdd{li255@email.sc.edu}
\emailAdd{creswick.rj@sc.edu}
\emailAdd{avignone@physics.sc.edu}
\emailAdd{wangyx@henu.edu.cn}

\abstract{In this paper we present a calculation of the sensitivity of the CUORE detector to the monoenergetic $14.4$ keV solar axions emitted by the M1 nuclear transition of$~^{57}$Fe in the Sun and detected by inverse coherent
Bragg-Primakoff conversion in single-crystal $TeO_2$ bolometers. The expected counting rate is calculated using density functional theory
for the electron charge density of $TeO_2$ and realistic background and energy resolution
of CUORE. Monte Carlo simulations for $5$ y $\times$ $741$ kg=$3705-$kg$\cdot$y of exposure are analyzed
using time correlation of individual events with the theoretical time-dependent counting rate. We find an expected model-independent limit on the product of the axion-photon coupling and the axion-nucleon coupling $g_{a\gamma\gamma}g_{aN}^{\text{eff}}<1.105\times 10^{-16}$ /GeV
for axion masses less than 500 eV with $95\%$ confidence level. 
}

\begin{document}
\maketitle
\flushbottom

\section{Introduction}
The strong CP problem in Quantum Chromodynamics (QCD), predicts the electric dipole moment of the neutron to be much larger than the observed upper limit \cite{NED}. Peccei and Quinn \cite{PQ1, PQ2} devised an elegent solution by introducing a new $U(1)_{PQ}$ global symmetry that is spontaneously broken at an energy scale $f_a$. A consequence of this $U(1)_{PQ}$ symmetry-breaking is that a new neutral spin-zero pseudoscalar particle (Nambu-Goldstone boson), the axion,  is generated \cite{Weinberg1978,Wilczek}. The axion acquries a mass through non-perturbation QCD effects. The ``standard axion'' with $f_a\approx f_{EW}=250$ GeV, where $f_{EW}$ is the electroweak scale, was quickly excluded by early searches \cite{Donnelly, Barshay, Barroso, Krauss}. Various models of ``invisible axions'' with $f_a>>f_{EW}$ have been proposed and recognized to be far-reaching because these axions can be a candidate for dark matter in the universe \cite{Preskill, Abbott, Fischler, Davis} and can be searched for by real experiments \cite{Bradley, Asztalos, Duffy}. Because the axion mass is inversely proportional to $f_a$, invisible axions are very light, very long-lived and very weakly coupled to photons, nucleons, electrons and quarks, which makes them difficult to detect directly. The two most widely cited models of invisible axions are the KSVZ(Kim, Shifman, Vainshtein and Zakharov) or hadronic axions \cite{KSVZ, KSVZ1} and the DFSZ (Dine, Fischler, Srednicki and Zhitnitskij) or GUT axions \cite{DFSZ, DFSZ1}. The main difference between the two models is that the KSVZ axions do not couple to ordinary leptons and quarks at tree-level.

Since axions, or more generally, axion-like particles (ALPs) can couple with electromagnetic fields or directly with leptons or quarks, the Sun could be an excellent axion emitter. Solar axions are generated by Primakoff conversion of photons, by Bremsstrahlung processes, by Compton scattering, by electron atomic recombination, by atomic deexcitation, and by nuclear M1 transitions. Axions produced in nuclear processes are monoenergetic because their energies correspond to the energy difference of a specific nuclear transition. These axions can be emitted and escape from the solar core due to the very weak interaction between the axion and matter. Searches for solar axions have been carried out with magnetic helioscopes \cite{CAST14p4, CASTBufferGas}, low temperature bolometers \cite{CUORE14p4} and thin foil nuclear targets \cite{Namba}. CUORE(Cryogenic Underground Observatory for Rare Events) \cite{CUORE, CUOREproposal} is designed to search for neutrinoless double beta decay($0\nu\beta\beta$) using a very low background low temperature bolometric detector. CUORE can also be used to search for dark matter WIMPs and solar axions. In this paper we calculate the expected conversion rate of $14.4$ keV solar axions produced in the M1 nuclear transition of$~^{57}$Fe and detected via the coherent inverse Primakoff process in $TeO_2$ single crystals. We present a calculation of the expected sensitivity of CUORE to the product of the coupling constant of axions to nucleons $g_{aN}^{\text{eff}}$ and the coupling constant of axions to photons $g_{a\gamma\gamma}$.

\section{Expected Counting Rates}

The coupling between the axion and electromagnetic field is described by the interaction Lagrangian
\begin{equation}
\mathscr{L}=-\frac{1}{4}g_{a\gamma\gamma}F^{\mu\nu}\tilde F_{\mu\nu}a=\frac{1}{M}\boldsymbol{E}\cdot\boldsymbol{B}a
\end{equation}
$F^{\mu\nu}$ is the electromagnetic field-strength tensor, $\tilde F_{\mu\nu}$ its dual, $\boldsymbol E$ and $\boldsymbol B$ the electric and magnetic fields, $a$ the axion field and $\frac{1}{M}=g_{a\gamma\gamma}$ the coupling constant.

\begin{figure}
    \subfloat[]{{\includegraphics[width=0.5\textwidth]{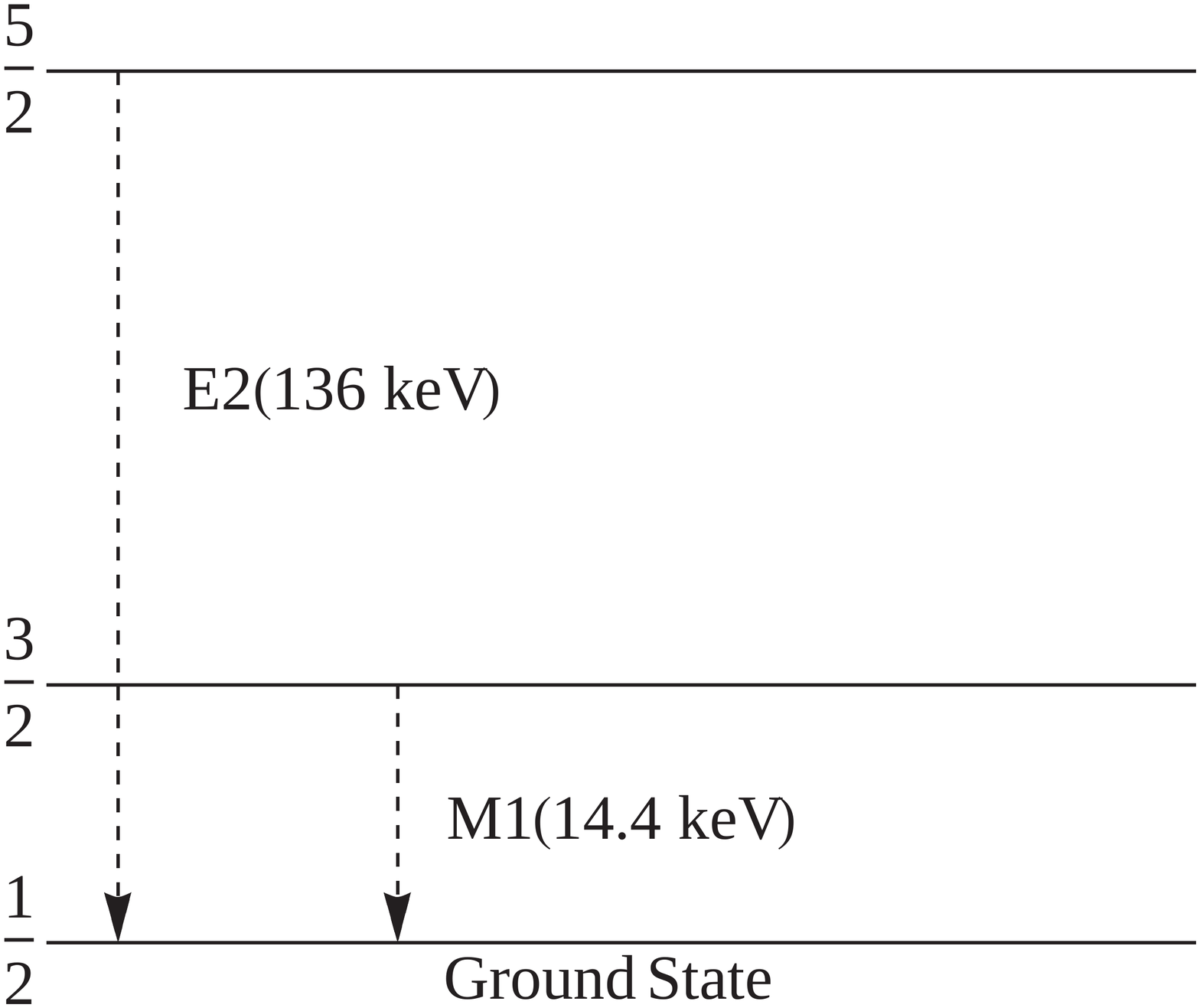} }}
    \subfloat[]{{\includegraphics[width=0.5\textwidth]{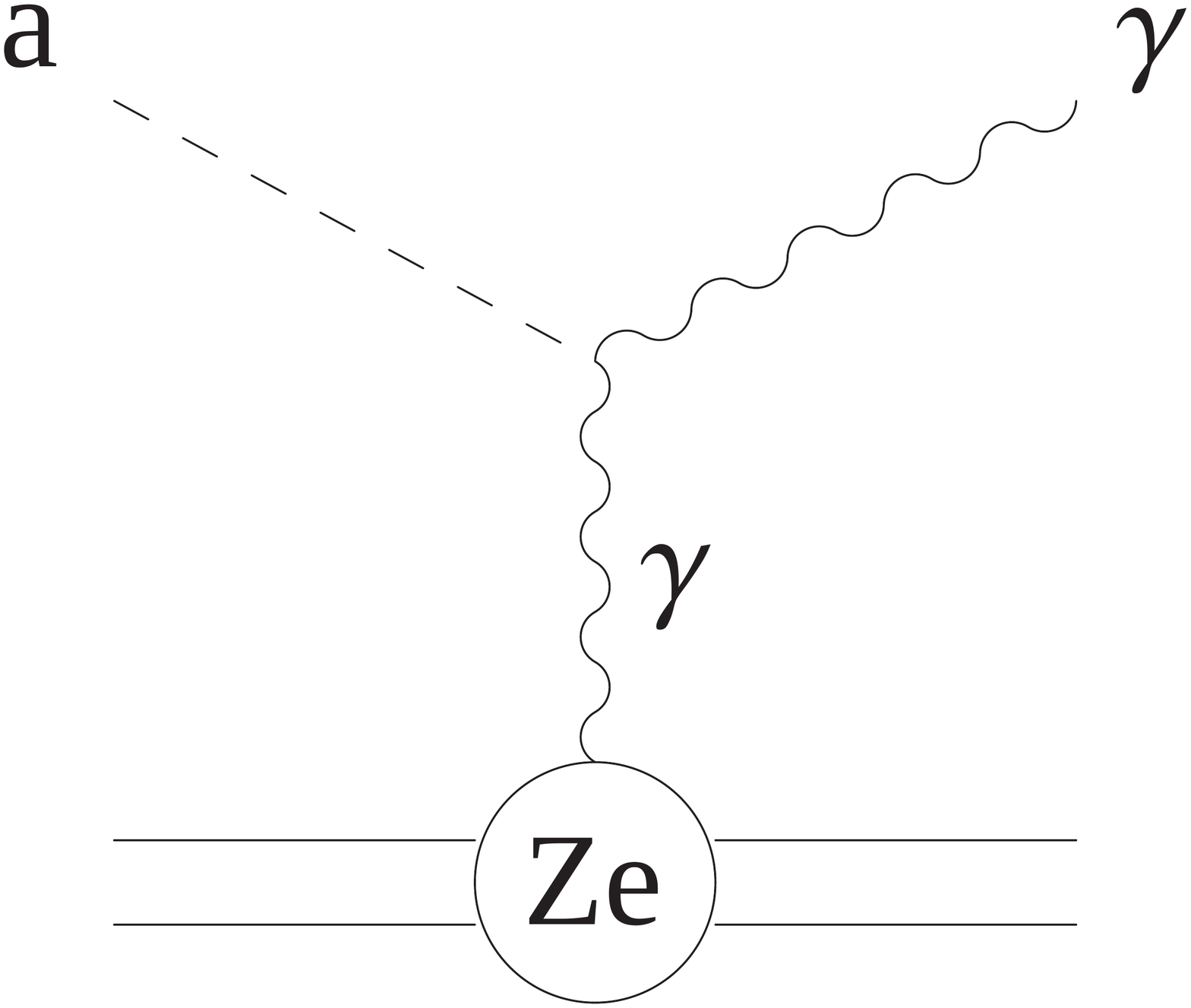} }}
   \caption{\label{fig:Primakoff}(a) An axion is emitted from the M1 nuclear transition from$~^{57}$Fe in the Sun: The energy difference between the first thermally excited state and the ground state of$~^{57}$Fe is $14.4$ keV. (b) An axion couples to a charge in the detector via a virtual photon in the Coulomb field of the crystal producing a photon by the inverse Primakoff effect \cite{TeO2}.}
\end{figure}

The matrix element for a conversion of an axion with energy $E_a$ and momentum $\boldsymbol p$ to a photon with polarization $\boldsymbol\epsilon$ energy $E_{\gamma}$ and momentum $\boldsymbol k$ was given earlier \cite{TeO2}:
\begin{equation}
\begin{split}
\mathcal M&=\langle\boldsymbol k\boldsymbol\epsilon;0|\mathcal H_{int}|0;\boldsymbol p\rangle \\
&=\frac{1}{V}\frac{\sqrt\alpha}{Mc^2}\frac{\hbar^3c^3}{2E_a^2}\frac{\boldsymbol\epsilon\cdot(\boldsymbol p\times\boldsymbol k)}{(\boldsymbol p-\boldsymbol k)^2}\tilde\rho(\boldsymbol p-\boldsymbol k)\delta(E_a-E_{\gamma}) 
\label{eq:matrixElement}
\end{split}
\end{equation}
where $V$ is the volume considered, $\alpha$ is the fine structure constant and $\tilde\rho$ is the Fourier transform of the charge density distribution which was calculated within density functional theory \cite{DFTtheorem, DFTequation} using WIEN2k \cite{WIEN2k},
\begin{equation}
\tilde\rho(\boldsymbol G)=\int_V\rho(\boldsymbol r)e^{-i\boldsymbol G\cdot\boldsymbol r}d^3r
\end{equation}
In a periodic lattice $\tilde\rho(\boldsymbol p-\boldsymbol k)$ vanishes unless $\boldsymbol p-\boldsymbol k=\boldsymbol G$, a reciprocal lattice vector, which means that the momentum transfer $\boldsymbol q$ must be equal to $\boldsymbol G$ \cite{TeO2}. 

The cross section as a function of the momentum of the axion for the conversion of an axion to a photon by the inverse Primakoff effect is \cite{TeO2}
\begin{equation}
\sigma_{a\gamma\gamma}=
m\hbar^3 c^3\frac{4\pi^2\alpha N_a}{\mu_cv_c}g_{a\gamma\gamma}^2\times\sum_{\boldsymbol G}|\tilde\rho_c(\boldsymbol G)|^2\frac{|\boldsymbol p\times\boldsymbol G|^2}{G^6}W_{\Delta}[E-E({\boldsymbol{\hat p},\boldsymbol G})]
\label{eq:crossSection}
\end{equation}
where m is the mass of the detector, $N_a$ Avogadro's constant, $\mu_c$ molar mass of the unit cell, $v_c$ is the volume of the conventional unit cell, $W_{\Delta}$ a Gaussian function with the same full width at half maximum as the detector and 
\begin{equation}
E(\boldsymbol{\hat p}, \boldsymbol G)=\hbar c\frac{G^2}{2\boldsymbol{\hat p}\cdot\boldsymbol G}
\label{eq:braggCondition}
\end{equation}
is the Bragg condition, which must be satisfied by the energy of the axion and direction to the Sun $\hat{\boldsymbol p}$ in order to have coherent conversion of axions to photons. It should be pointed out that in Eq.~\eqref{eq:crossSection} we have included factors of $\hbar$ and $c$ explicitly. If the axion has a mass $m_a$ the Bragg condition in Eq.~\eqref{eq:braggCondition} is modified to
\begin{equation}
E(\boldsymbol{\hat p}, \boldsymbol G)=\sqrt{\hbar^2c^2\left(\frac{G^2-\frac{m_a^2c^2}{\hbar^2}}{2\boldsymbol{\hat p}\cdot\boldsymbol G}\right)^2+m_a^2c^4}
\end{equation}
For axion masses less than $500$ eV the shift in the Bragg peaks and the flux from the Sun are only a few percent. For axion masses approaching $1$ keV these effects become more pronounced, so we place an arbitrary and conservative exclusion limit on axions with masses less than $500$ eV. CUORE will have a characteristic low background counting rate of $2$ cpd/keV/kg at $25$ keV and a low-energy resolution $\Delta=$FWHM=$0.73$ keV at $4.7$ keV \cite{CUOREresolution}.

The natural abundance of the stable isotope of$~^{57}$Fe in the core of the Sun is $2.2\%$ and the mass fraction of$~^{57}$Fe $2.8\times 10^{-5}$. The first excited state of$~^{57}$Fe is at $14.4$ keV and can be thermally excited in the interior of the sun ($kT\approx 1.3$ keV). The excited nucleus can relax to the ground state by emitting a $14.4$ keV photon or an internal conversion electron. Emission of an axion from the first excited state is also possible. The coupling between the axion and nucleons is described by the interaction Lagrangian \cite{CAST14p4}
\begin{equation}
{\mathscr L}_{aN}=-ia\bar\psi_N\gamma_5\left(g_{aN}^0+g_{aN}^3\tau_3\right)\psi_N
\end{equation}
where $\psi_N$ is the nucleon doublet, $\tau_3$ is the Pauli matrix, and $g_{aN}^0$ and $g_{aN}^3$ are the isoscalar and isovector axion-nucleon coupling constants, respectively. The first search for monoenergetic$~^{57}$Fe solar axions was proposed by Moriyama \cite{Moriyama} and the flux was calculated by Haxton and Lee \cite{Haxton}
\begin{equation}
\Phi_{Fe}=4.56\times 10^{27}(g_{aN}^{\text{eff}})^2m^{-2}s^{-1}
\label{eq:FeFlux}
\end{equation}
where $g_{aN}^{\text{eff}}\equiv-1.19g_{aN}^0+g_{aN}^3$ is the effective axion-nucleon coupling constant \cite{CAST14p4,CUORE14p4}.
The total conversion rate of axions from the M{1} transition of $~^{57}$Fe to photons by the inverse Primakoff process as a function of the momentum of the axion is
\begin{equation}
\begin{split}
\frac{d\dot N}{dE}&=\Phi_{Fe}\times\sigma_{a\gamma\gamma}\\
&=4.56\times 10^{27}\times(g_{aN}^{\text{eff}}g_{a\gamma\gamma})^2m\hbar c\frac{4\pi^2\alpha N_a}{\mu_cv_c}\times\sum_{\boldsymbol G}|\tilde\rho_c(\boldsymbol G)|^2E^2\frac{|\boldsymbol{\hat p}\times\boldsymbol G|^2}{G^6}W_{\Delta}[E-E({\boldsymbol{\hat p},\boldsymbol G})]
\label{eq:conversionRate}
\end{split}
\end{equation}
Integrating $d\dot N/dE$ over an energy range $14.1\leq E_a\leq 14.7$ keV given by the resolution of the bolometer, $\Delta E=0.6$ keV centered at $14.4$ keV gives the time-dependent counting rate
\begin{equation}
R(t)=\int_{E'}^{E'+\Delta E}\frac{d\dot N}{dE}(t, E'')dE''
\label{eq:axioncountingrate}
\end{equation}

\begin{figure}[h]
\centering
\includegraphics[scale=0.4]{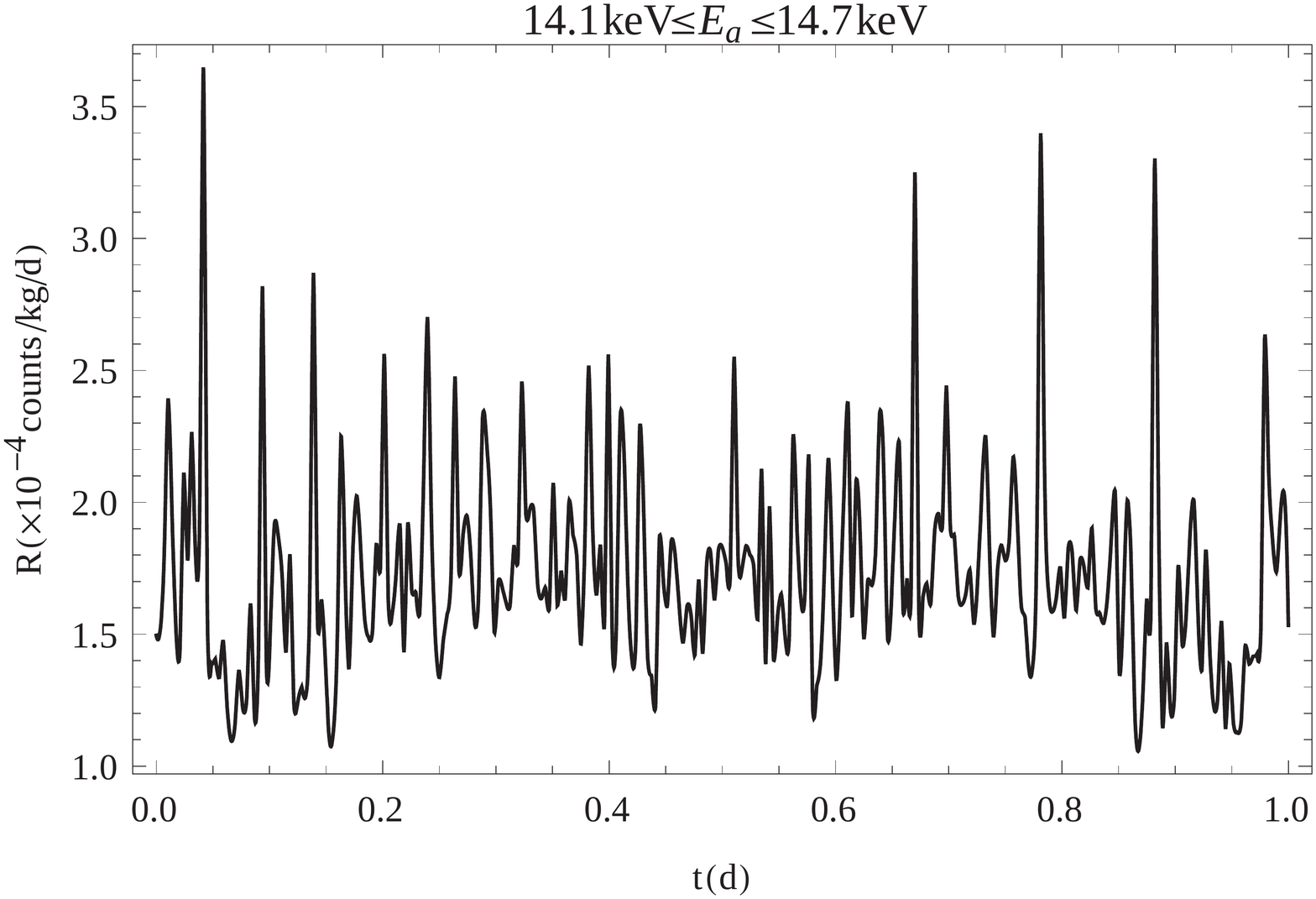}
\caption{Expected counting rates $R(t)$ of photons produced by the inverse Primakoff conversion of $14.4$ keV solar axions in the CUORE detector for $g_{a\gamma\gamma}g_{aN}^{\text{eff}}=1.105\times 10^{-16}$ $\text{GeV}^{-1}$. The counting rate was calculated for $1$ day using Eq.~\eqref{eq:conversionRate} and Eq.~\eqref{eq:axioncountingrate}.}
\label{fig:countingrate}
\end{figure}
Figure~\ref{fig:countingrate} shows the expected counting rate of the $14.4$ keV solar axions as a function of time over a single day with the flux given in Eq.~\eqref{eq:FeFlux} and $g_{a\gamma\gamma}g_{aN}^{\text{eff}}=1.105\times 10^{-16}$ $\text{GeV}^{-1}$.

\section{Monte Carlo Simulation}
We use the time correlation of individual events with the theoretical time-dependent counting rate to calculate the sensitivity of the CUORE detector to $g_{a\gamma\gamma}g_{aN}^{\text{eff}}$\cite{TeO2}. 
\begin{figure}[h]
\centering
\includegraphics[scale=0.4]{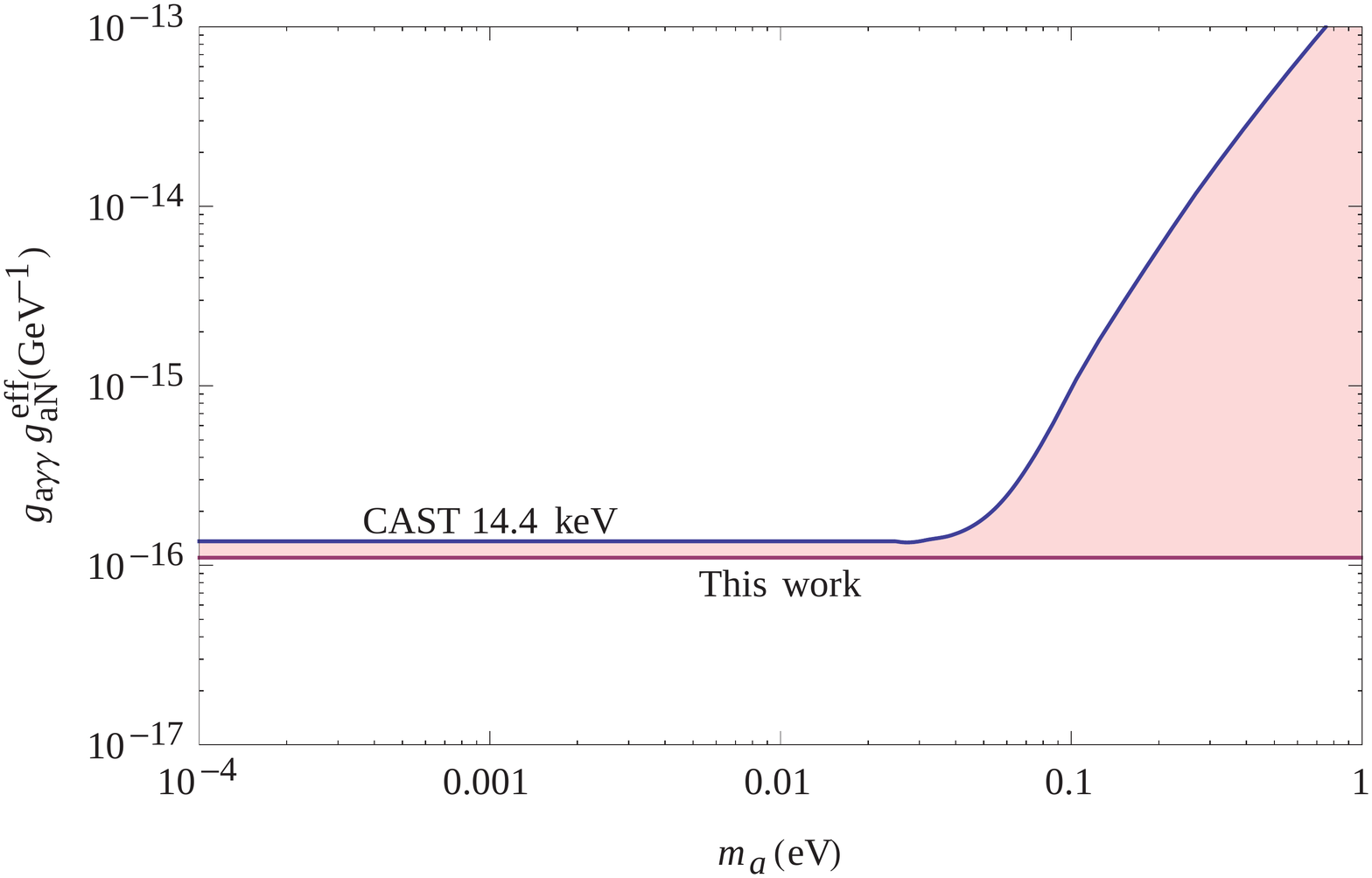}
\caption{Exclusion limits on the $g_{a\gamma\gamma}g_{aN}-m_a$ plane. The shaded region is the potential constraint for axion masses less than $500$ eV by CUORE detector with five years of data.}
\label{fig:garrganVSma}
\end{figure}
The Monte Carlo simulation for $741-$kg$\cdot$y can set a model-independent upper bound for the product of the axion-photon and the axion-nucleon coupling constants $g_{a\gamma\gamma}g_{aN}^{\text{eff}}<2.47\times10^{-16}$ $\text{GeV}^{-1}$. To illustrate the resolving power of the time correlation method for the $14.4$ keV solar axions, there are approximately 300 events due to axion conversion and $1.05\times10^5$ background events in one year with $g_{a\gamma\gamma}g_{aN}^{\text{eff}}=2.47\times10^{-16}$ $\text{GeV}^{-1}$. With five years of data, CUORE can set an upper bound of $g_{a\gamma\gamma}g_{aN}^{\text{eff}}<1.105\times10^{-16}$ $\text{GeV}^{-1}$, which is slightly better than the current bound set by CAST for $m_a<0.03$ eV, as shown in Figure~\ref{fig:garrganVSma}. For ten years simulation, the upper bound can be reached to $g_{a\gamma\gamma}g_{aN}^{\text{eff}}<0.781\times10^{-16}$ $\text{GeV}^{-1}$.

\section{Conclusions}

\begin{figure}[h!]
\centering
\includegraphics[scale=0.4]{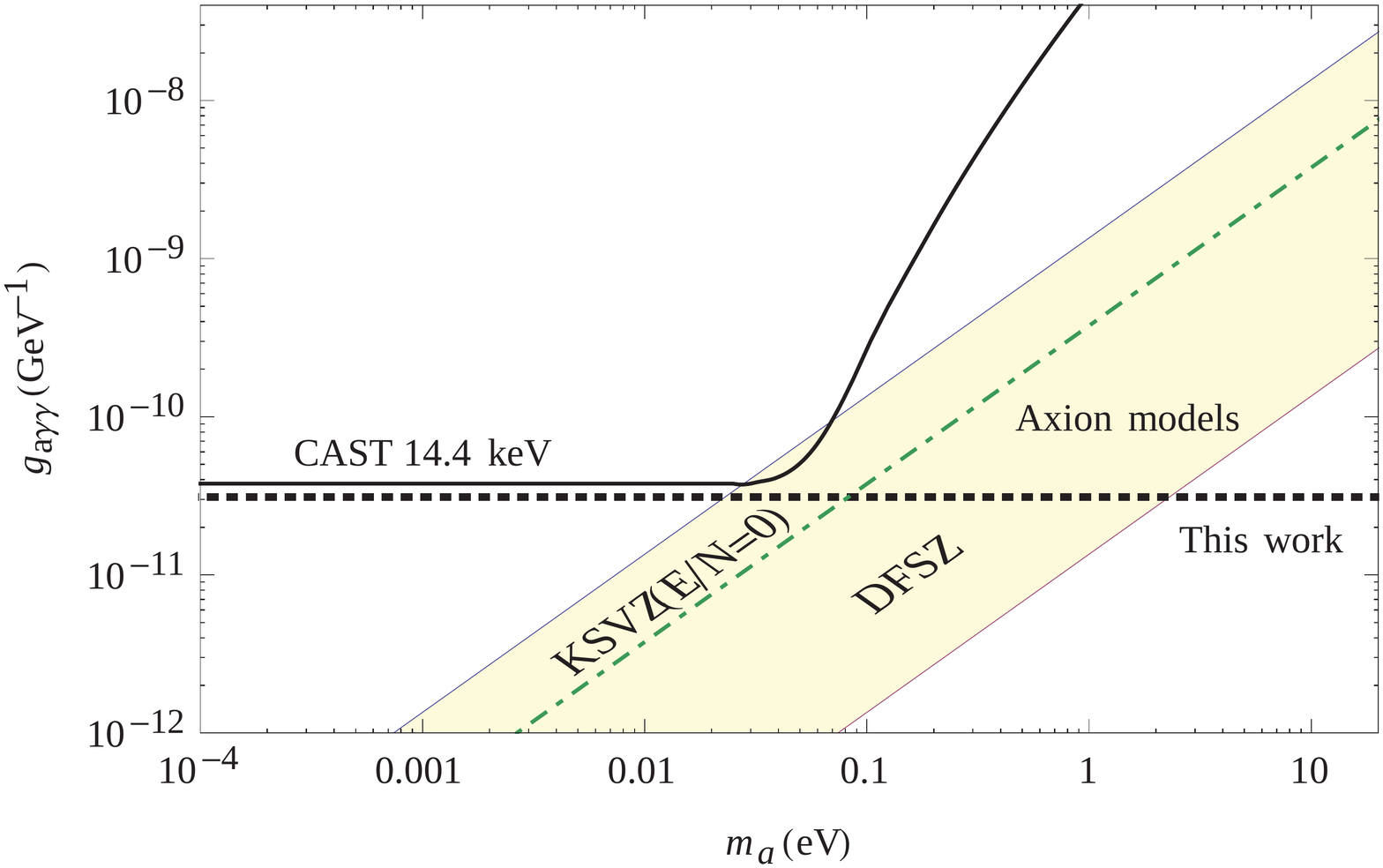}
\caption{Exclusion limits on the $g_{a\gamma\gamma}-m_a$ plane. The dotted line is a relative limit on the $g_{a\gamma\gamma}$ coupling constant with $g_{aN}=3.6\times 10^{-6}$. The value $g_{aN}=3.6\times 10^{-6}$, used in Ref \cite{CAST14p4}, was used here in order to make a direct comparison to the sensitivity achieved in the CAST experiment.}
\label{fig:garrVSma}
\end{figure}

Figure~\ref{fig:garrVSma} shows the excluded region of the $g_{a\gamma\gamma}-m_a$ plane achieved by CAST assuming $g_{aN}=3.6\times 10^{-6}$, which is set by the requirement that the$~^{57}$Fe solar axion luminosity should be less than $10\%$ of the solar photon luminosity($L_a<0.1 L_{\odot}$)\cite{Luminosity}. The dotted line is a bound for $g_{a\gamma\gamma}$ that could be set by CUORE under the same assumption. Our simulation shows that CUORE could eliminate a substantial part of the model space not yet touched by other experiments.

\acknowledgments

This work was supported by the US National Science Foundation Grant PHY-1307204.

\end{document}